\documentclass[10pt,letterpaper]{article}
\usepackage{opex3}

\begin{document}

\title{All-optical link for direct comparison \\of distant optical clocks}

\author{Miho Fujieda$^{1}$, Motohiro Kumagai$^{1}$, Shigeo Nagano$^{1}$,
\\Atsushi Yamaguchi$^{1,2}$, Hidekazu Hachisu$^{1,2}$, and Tetsuya Ido$^{1,2}$}

\address{$^{1}$National Institute of Information and Communications
Technology, \\4-2-1, Nukui-kitamachi, Koganei, Tokyo, 184-8795 Japan\\
$^{2}$CREST, Japan Science and Technology Agency, \\4-1-8, Honcho,
Kawaguchi, Saitama, 332-0012 Japan}

\email{miho@nict.go.jp} 



\begin{abstract}
We developed an all-optical link system for making remote comparisons 
of two distant ultra-stable optical clocks.
An optical carrier transfer system based on a fiber interferometer 
was employed to compensate the phase noise accumulated during the 
propagation through a fiber link. Transfer stabilities of 
$2\times10^{-15}$ at 1 second and $4\times10^{-18}$ at 1000 seconds 
were achieved in a 90-km link. 
An active polarization control system was additionally introduced 
to maintain the transmitted light in an adequate polarization, 
and consequently, a stable and reliable comparison was accomplished. 
The instabilities of the all-optical link system, including 
those of the erbium doped fiber amplifiers (EDFAs) which are free from 
phase-noise compensation, 
 were below 
$2\times10^{-15}$ at 1 second and $7\times10^{-17}$ at 1000 seconds. 
The system was available for the direct comparison 
of two distant $^{87}$Sr lattice clocks via an urban fiber link of 60 km. 
This technique will be essential 
for the measuring the reproducibility of optical
 frequency standards. 
\end{abstract}



\section{Introduction}

Clock comparisons are essential to confirm the 
reproducibility of frequency standards. 
For the past twenty years, it has been possible to remotely calibrate 
the frequencies of microwave atomic clocks by using 
satellite links such as GPS or two-way satellite time and
frequency transfer (TWSTFT). One of the applications of 
these techniques is the network of
international atomic time (TAI) links established 
by Bureau International des Poids et Mesures (BIPM), and 
it allows us to know time difference 
between clocks separated by 
some distance \cite{Panfilo2010}. 
It is provided by BIPM that 
today's typical satellite links using GPS precise point positioning
 and TWSTFT have 
frequency stabilities of \(1\times10^{-15}\) and
\(2\times10^{-15}\), respectively,  for an averaging time of five days. 
On the other hand, stabilities of optical frequency standards 
have rapidly increased, 
reaching \(1\times10^{-16}\) for an averaging time
of 1000 seconds \cite{Chou2010}.This progress has led to the need for 
a technique to remotely 
compare frequencies without degrading their stabilities. 
However, the frequency stabilities of satellite-based links are 
apparently insufficient which spoil  
the benefit of optical standards. 
Instead, 
transferring a stable optical frequency over optical fibers 
is regarded as a strong candidate for making direct comparisons 
of highly stable optical clocks and, it 
has been studied extensively \cite{Foreman2007,Williams2008,Musha2008}. 
Successful experiments have been performed 
over a short distance link \cite{Ma1994} and with a spooled fiber
\cite{Newbury2007}. Other studies include ones on transfer of a clock 
signal together with data traffic 
\cite{Kefelian2009, Lopez2010}, using a high gain amplifier for
a long-haul link up to 480 km \cite{Terra2010}, 
and developing a repeater for cascaded systems \cite{Lopez2010}. 
Moreover, optical carrier transfer techniques have been 
developed to observe the relative frequency stability of optical
clocks and cavity stabilized lasers \cite{Ludlow2008, Terra2009, Pape2010}. 
Stable optical frequency transfer would be useful 
not only for making frequency comparisons, but also for 
optical coherent communications.

We developed an all-optical link system that consists of 
Ti:sapphire (Ti:S) frequency combs, nonlinear crystals for frequency doubling,
fiber amplifiers, a 1.5 $\mu$m stable laser, an optical carrier transfer system, and an active polarization control system. 
The system can perform reliable
measurements and can operate for a long time free  
from large polarization variations caused by the long-haul fiber link. 
Its feasibility was confirmed in an experiment that  
directly compared two \(^{87}\)Sr clocks 
located at the National Institute of Information and Communications
Technology (NICT) and the University of Tokyo (UT). 
In this paper the details of the system and its performance are
described with the result of direct comparison of distant
optical clocks.

\section{Overall system}

\begin{figure}[hbt]
\centering\includegraphics[width=12cm]{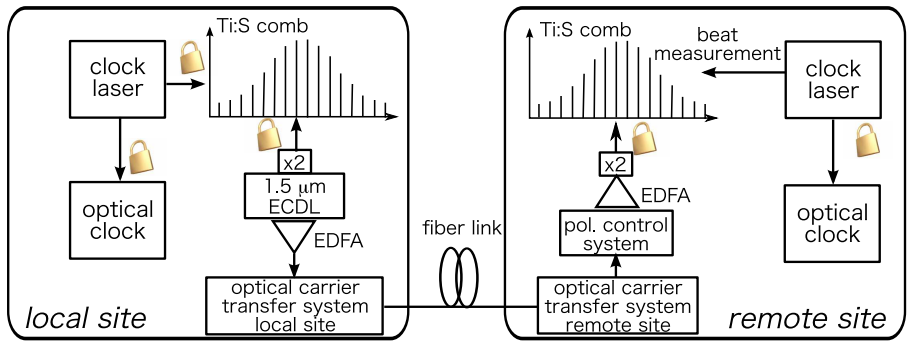}
\caption{Overall system of the all-optical link. The clock
 signal at the local site is frequency-converted into the 1.5 $\mu$m light
 and transferred 
to the remote site via an optical fiber link. 
The optical carrier transfer system compensates the phase noise so 
 that the clock signal is faithfully transferred 
to the remote site. 
The transferred light is converted into a visible wavelength to be compared
 with another clock signal at the remote site. ECDL: external cavity diode laser, x2: PPLN frequency doubler.}
\label{fig:overall}
\end{figure}

Fig. \ref{fig:overall} is a diagram of the all-optical link system. 
A stable optical frequency is sent from the local site to the remote site 
through an optical fiber link. 
The telecom wavelength at 1.5 \(\mu\)m 
is adequate for transmissions 
over optical fiber links from the perspective of low optical loss, while 
most of the optical clock transitions 
lie in the range of visible light. Hence, there is a need for 
frequency conversion between the two bands,  
and the system uses two Ti:S frequency combs 
and two periodically poled lithium
niobate (PPLN) for this purpose. 
The clock lasers at the both sites 
are locked to an optical transition of 
atoms or ions. 
The Ti:S frequency comb is phase-locked to the clock laser at the local
site. 
An external-cavity diode laser (ECDL) operating at 1.5 $\mu$m is 
phase-locked to the frequency comb through its 
second harmonic
generation (SHG) light generated by the PPLN. 
The light emitted from the ECDL is amplified by a booster EDFA 
and an optical carrier transfer system, which is described below, 
delivers the light to the remote site. 
At the remote site, the transferred light is amplified 
by an EDFA in order to feed enough power into the PPLN there. 
The state of polarization (SOP) is stabilized by the 
polarization control
system and the light is frequency doubled. 
The Ti:S frequency comb at the remote site is phase-locked to the SHG light. 
The beat signal between the clock laser and the
nearest comb component represents the differential frequency and 
the relative stability of the two optical clocks.

\section{Subsystems}
\subsection{Optical carrier transfer system}

\begin{figure}[hbt]
\centering\includegraphics[width=10cm]{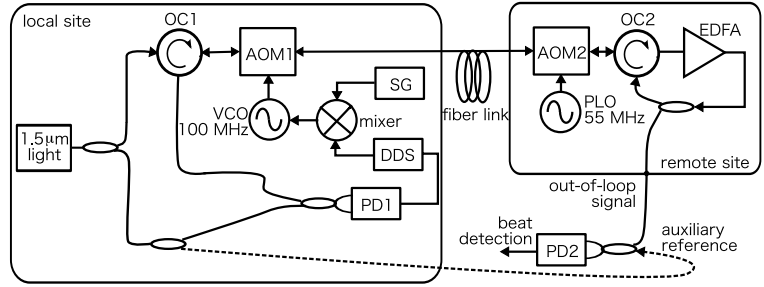}
\caption{Schematic diagram of the optical carrier transfer system. The system
 consists of local and remote parts. To evaluate the system
 performance, both parts are initially located 
at the same place. The beat
 signal between the output of the remote site and auxiliary reference is
 used for the evaluation. OC: optical
 circulator, AOM: acousto-optic modulator, VCO: voltage controlled
 oscillator, SG: signal generator, DDS: direct digital synthesizer, PD:
 photo detector, PLO: phase locked oscillator, EDFA: erbium doped fiber
 amplifier. }
\label{fig:carrier_system}
\end{figure}

Fig. \ref{fig:carrier_system} shows a schematic diagram of the optical carrier
transfer system with which we tested the fiber noise
cancellation capability. 
To compensate the accumulated phase noise during the transmission, 
we developed a noise cancellation system based on a fiber 
interferometer. 
The concept is similar to the scheme 
demonstrated by Ma et al. \cite{Ma1994}. 
A narrow-bandwidth 1.5 $\mu$m light to be transferred is coupled to 
a fiber-pigtailed acousto-optic modulator 1 (AOM1) driven by a 100-MHz 
voltage controlled oscillator (VCO) via an optical circulator 1 (OC1) 
and transmitted to the remote site through a long-haul link. 
At the remote site, the transmitted light is connected to the 
second fiber-pigtailed AOM 2 (AOM2) driven by a 55-MHz stable 
reference and 
amplified by a uni-directional EDFA. 
Part of the amplified light is served to a user (denoted as the 
``out-of-loop signal''), and the rest is sent back to the local site
through optical circulator 2 (OC2) and  AOM2. 
AOM2 discriminates the returned signal from 
the stray reflections at connectors and splices. 
One concern in using the uni-directional EDFA
is that only half of the noise induced by it is compensated by
the system. As described later, we found that 
the remaining half of the noise 
does not limit the performance of our system.
Bi-directional EDFAs and Faraday rotator mirrors (FRMs) 
are normally 
used to compensate optical loss and to reflect part of the
transmitted light back to the local site. 
Indeed, the combination of the bi-directional EDFA and the FRM worked 
properly in performance checks using a spooled fiber or a 
dedicated fiber link. 
However, in the case applying a bi-directional EDFA to the fiber link 
from NICT to UT, 
unwanted back-reflections from many SC/PC connectors at the remote site 
induced an 
excessive amount of input to the bi-directional EDFA, and this 
resulted in saturation of the output. 
The returned light to the local site is 
coupled to a photo detector 1 (PD1) via AOM1 and 
OC1. AOM1 and AOM2 used -1$^{st}$ and +1$^{st}$
diffraction, respectively. The frequency of the return light which 
passes AOM1 and AOM2 twice is shifted by -90 MHz. 
The polarization change in the
traveling light is an important issue in the long-haul link. 
Retro-reflection by the FRM makes it possible to maintain the difference 
of the state of polarization (SOP) between the reference and return
light. In our case free from the FRM, the return light is differentially
detected by a balanced photo detector to moderate the amplitude
variation of the 90-MHz beat signal. The obtained beat signal is 
divided-by-50 to 1.8 MHz by a direct digital synthesizer (DDS).  
The DDS works not only to expand the capture range of the phase lock 
for handling of a huge amount of phase noise 
but also to make the amplitude of the beat signal uniform. 
The resulting divided signal is mixed with a 1.8-MHz stable reference 
linked to a hydrogen maser. The mixer output is amplified,
filtered, and fed back to the VCO. 
Thus, the phase of the VCO is adjusted to compensate the fiber induced noise. 

\subsection{JGN2plus optical fiber link}

\begin{figure}[hbt]
\centering\includegraphics[width=6cm]{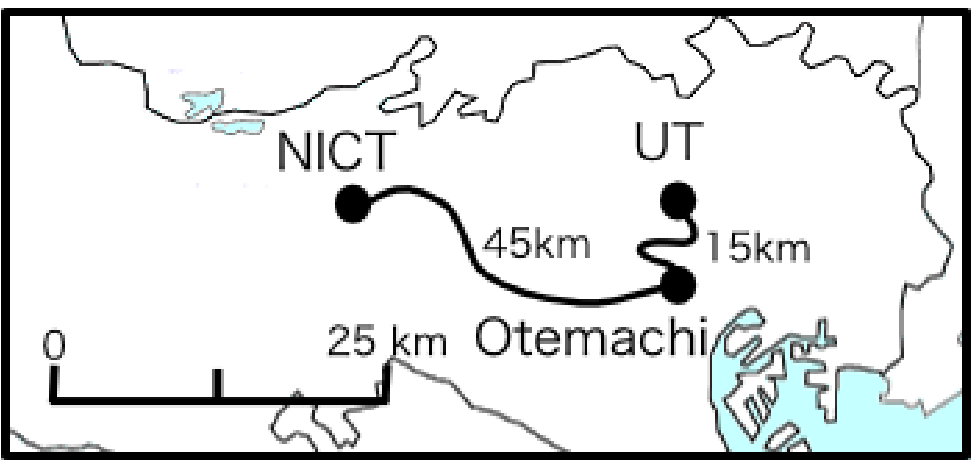}
\hspace{8mm}
\centering\includegraphics[width=5cm]{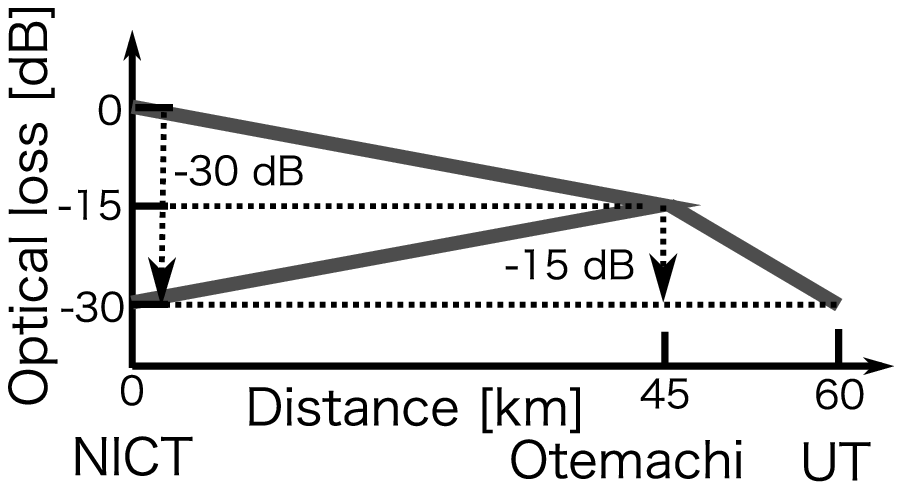}
\caption{Schematic diagram of the optical fiber link in Tokyo (left) and the optical
 losses (right).}
\label{fig:fiber_link}
\end{figure}

\begin{figure}[htb]
\vspace*{-20mm}
\hspace*{-10mm}
\centering\includegraphics[width=14cm]{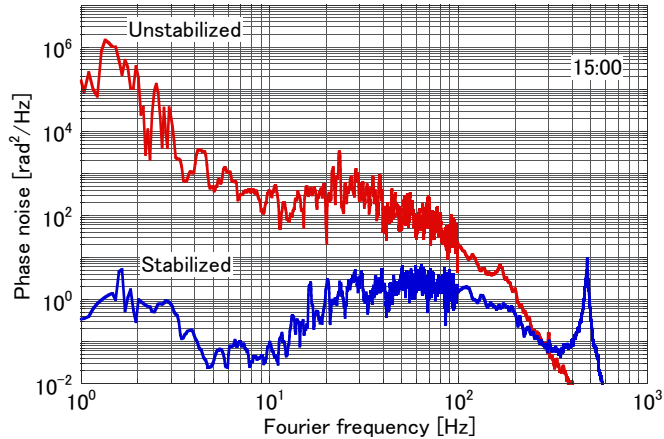}
\vspace*{-27mm}
\caption{Phase noises of the out-of-loop beat in the 90-km unstabilized (red)
 and stabilized (blue) links. The difference of 56 dB at 1 Hz agrees
 with the theoretical limit of phase noise suppression. Thus, the
 optical carrier system works properly.}
\label{fig:phasenoise_ft-system}
\end{figure}

\begin{figure}[hbt]
\centering\includegraphics[width=10cm]{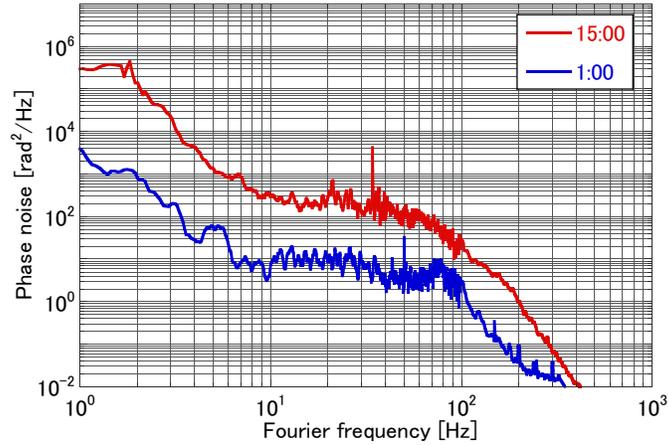}
\vspace*{-8mm}
\caption{Phase noise of the 60-km unstabilized link between NICT and
 UT. 
The data were obtained at 1:00 and 15:00. The
 difference between the daytime and nighttime results is one
 or two orders of magnitude. }
\label{fig:phasenoise_ut}
\end{figure}

NICT operates an optical fiber network test bed named JGN2plus, as 
infrastructure for research and development 
on information and communications technology
\cite{JGN2+}. 
Part of the test bed connects NICT and Otemachi (a business district
in Tokyo) 
45 km away by interconnecting several sections of 
single-mode dark fiber. 
To characterize the phase noise cancellation system, 
we used two parallel links, and joined them at Otemachi 
so that the local and remote sites of a 90-km link are both 
at NICT.  
Fig. \ref{fig:fiber_link} shows a schematic diagram of the link. 
The optical loss is -30 dB for the round-trip link. 
Fig. \ref{fig:phasenoise_ft-system} depicts the power spectral density
of the phase noise imposed on the  
1.5 $\mu$m light traveling in the round-trip 90-km optical fiber link. 
This phase noise is much larger than the noises in 
other optical fiber links \cite{Kumagai2009, Fujieda2010}.
For example, Terra et al.
\cite{Terra2010} reported phase noise of about 200 rad\(^{2}\)/Hz at
a Fourier frequency of 1 Hz in a 480-km link, 
which is 30 dB less than that in the JGN2plus link. 
The large amount of noise is probably 
due to almost 
half of the link between NICT and Otemachi being buried 
along a subway line and about one third being 
wired in the air. 
This speculation is born out by 
the fact that the phase noise increases on windy days 
and decreases at nighttime 
when the subway is out of service. 
In addition, the SOP of the transferred light changes dynamically 
during the transmission because of deformation of the 
optical fiber core due to temperature and 
pressure variations. 
Stable transfer of the frequency standard signal 
requires not only compensation of a huge amount of 
phase noise but also active polarization control.

NICT was connected to UT by a fiber link with a 
total length of 60 km by extending the optical fiber 
link from Otemachi to UT for 15 km. 
The physical distance between the two laboratories is 24 km. 
The optical loss in the additional 15 km is -15 dB. 
This relatively large optical loss is due to the 
fiber connectors in the multiply-connected link. 
Fig. \ref{fig:phasenoise_ut} shows the power spectral density of 
the phase noises accumulated on the 60-km NICT-UT 
link. They were computed from measurements on 
light traveling in the free-running 
round-trip NICT-UT-NICT link of 120 km. 
The level is almost the same as 
that of the 45-km link, indicating that the path between 
NICT and Otemachi predominantly induces the phase noise. 
We evaluated the system in the 90-km link and used this 
evaluation to estimate the upper limit of 
the system performance for the 60-km link.

\subsection{Performance of the optical carrier transfer system}

\begin{figure}[hbt]
\centering\includegraphics[width=11cm]{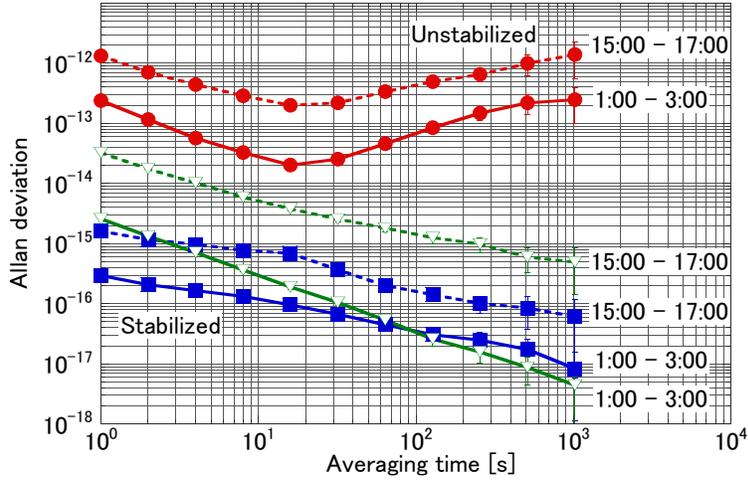}
\vspace*{-12mm}
\caption{Frequency stabilities of the transferred signal in the
 unstabilized 
(red) and stabilized (blue
 \& green) links. The curves in red and green colors were measured by a
 $\Pi$-type frequency counter. The blue curves were measured by a
 $\Lambda$-type frequency counter. 
The frequency stabilities shown in the dashed and solid curves were
 measured during the daytime (15:00-17:00) and around 
midnight (1:00-3:00), respectively.}
\label{fig:adev_ft-system}
\end{figure}

The out-of-loop signal is evaluated by mixing with the reference light, 
where a 1538-nm fiber laser is used as a light source. 
The phase noise of the out-of-loop signal 
was measured as shown 
in Fig. \ref{fig:phasenoise_ft-system}. 
The blue and red curves show results with 
and without link stabilization, respectively. 
These results were obtained 
at 15:00 in the daytime. 
The delay of propagation gives limitations of the servo loop
bandwidth, which constrains the loop gain of the servo 
to a finite value. 
The theoretical maximum of noise suppression is 
represented as \(1/3(2\pi f\tau)^{2}\) \cite{Williams2008}, 
where \(f\) is the Fourier frequency and \(\tau\) is
the one-way propagating time.
The maximum cancellation is 56 dB in 
90-km links at a
Fourier frequency of 1 Hz. 
As seen in Fig. \ref{fig:phasenoise_ft-system}, the suppression ratio reaches 
the theoretical limit. 
The bump at 470 Hz found in the stabilized link shows 
the servo bandwidth for phase locking. 
Fig. \ref{fig:adev_ft-system} depicts the 
frequency stability of the
out-of-loop signal. 
It is known that $\Lambda$-type frequency counters with a 
dead time cannot produce the appropriate 
slope of $\tau^{-1}$ in the Allan deviation 
when a signal with white phase noise is 
counted \cite{Lesage1983}. 
We show the results obtained with 
a $\Pi$-type counter (green) 
as well as with a $\Lambda$-type counter (blue).  
The dashed and solid lines respectively show the stabilities
measured from 15:00 to 17:00 in the daytime, 
and from 1:00 to 3:00 at 
nighttime. 
There is an order of difference between the short-term stabilities 
obtained by the $\Pi$-type and $\Lambda$-type counters. 
This discrepancy might be attributed to a measurement bandwidth 
of each frequency counter. 
At nighttime, the transfer stability measured by a 
$\Pi$-type counter has the $\tau^{-1}$ slope, 
resulting in a stability of $4\times10^{-18}$ 
for an averaging time of 1000 seconds. 
On the other hand, the transfer stability in the daytime has 
$\tau^{-1/2}$ slope, although it was measured 
with a $\Pi$-type counter. 
This results from the noise component of the unstabilized link in the
daytime. As shown in Fig. \ref{fig:phasenoise_ft-system}, 
the phase noise of the unstabilized link has the 
dependence close to $1/f^{4}$ 
between 1 Hz and 10 Hz. Since fiber noise cancellation 
follows an $f^{2}$ law \cite{Williams2008}, the slope of the 
residual phase noise during the 
daytime is approximately $1/f^{2}$ in the
stabilized link. Consequently, the stabilized link shows  
a $\tau^{-1/2}$ trend like a signal dominated by white FM noise. 
At nighttime, the fiber noise changes form to 
a $1/f^{3}$ dependence, 
resulting in a $\tau^{-1}$ slope. 
This large difference between daytime and nighttime 
is attributed to the activity of the subway in
Tokyo which is out of service at midnight. 
In spite of the huge phase noise in the JGN2plus link, 
the transfer stability reached the $10^{-18}$ level at 
nighttime, 
which is much more stable than a conventional 
satellite-based link. 
Additionally, the observed mean deviation of the transferred carrier
frequency was 
\(\Delta\nu=(-2.0\ \pm\ 0.4)\) mHz for the measurement from 1:00 to 3:00 at
nighttime, which was within the statistical
uncertainty.

\subsection{Bridge between clock transition and telecom wavelength}
\subsubsection{Active polarization control and reliable measurement}

\begin{figure}[hbt]
\centering\includegraphics[width=8cm]{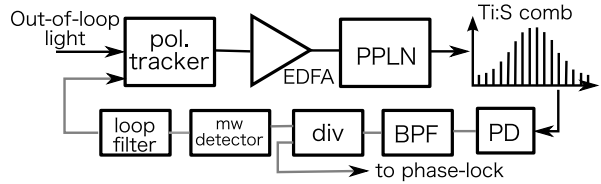}
\caption{Schematic diagram of the polarization control system. The
 polarization variation is detected by the beat intensity between the
 SHG light and the relevant Ti:S comb component. 
The rotation is compensated at the polarization tracker (pol. tracker). 
PPLN: periodically poled lithium niobate, 
 div: divider, 
 BPF: band-pass filter, mw: microwave.}
\label{fig:tracker_system}
\end{figure}

\begin{figure}[htb]
\centering\includegraphics[width=12cm]{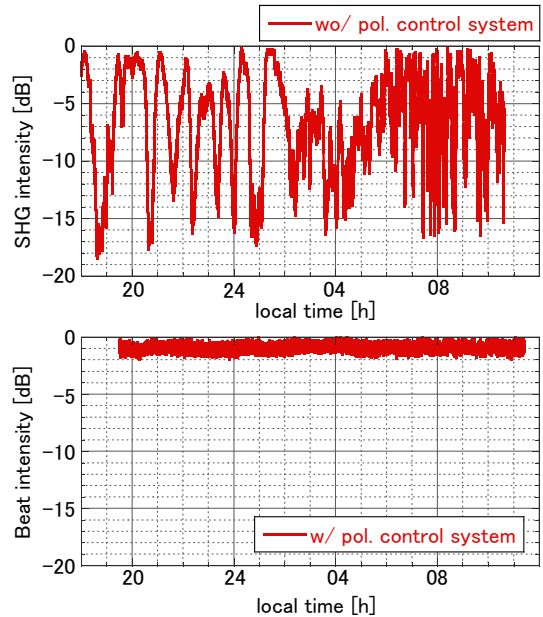}
\caption{Up: Intensity variation of the SHG light without the polarization
 control system. Down: Intensity variation of the beat signal with the
 polarization control system. Both results were obtained in the 90-km link.}
\label{fig:tracker_on_off}
\end{figure}

The conversion efficiency of PPLN is quite sensitive 
to the SOP of the input light. 
Hence, active polarization control is essential for reliable 
measurements without interruption. 
Our automatic polarization control system based on  
a commercial polarization tracker is designed to distinguish 
the signal in the input light from false components caused 
by reflections from return light 
at connectors and splices. 
Fig. \ref{fig:tracker_system} shows 
the schematic diagram of the polarization control system. 
The out-of-loop signal is coupled to the PPLN 
through the polarization tracker and the EDFA. 
The SOP variation due to the EDFA is also 
compensated in this system. 
The intensity of the correct component of the SHG light 
is identified as the RF amplitude of 
the specific beat frequency against 
the Ti:S frequency comb. 
The output SHG light is down-converted to 
the RF domain by the beat detection, where the beat note with 
the desired component 
is selectively obtained by a narrow band-pass filter. 
The intensity is detected with a microwave detector 
(a planar doped barrier detector). 
The commercial polarization tracker controls the SOP of the 
light by driving a fiber squeezer so that the deviation of the feedback signal is reduced. 
The tracking speed is 47$\pi$/s, and 
the typical SOP recovery time is 0.7 ms, 
which is fast enough to cancel out the SOP 
variation in the dedicated link. 
Fig. \ref{fig:tracker_on_off} shows the behaviors of the intensity fluctuations of 
the beat signal and SHG light with and without 
the polarization control system
in the 90-km optical fiber link. 
The results show that the 
system significantly reduced the intensity fluctuation due
to the polarization variation and kept the intensity fluctuation of the 
beat signal within
2 dB for 16 hours. This novel system is thus capable of 
continuous operation without manual 
polarization adjustments. 
However, a small intensity fluctuation remained 
even though the SOP is effectively stabilized 
by our polarization control system. 
If an intensity fluctuating signal were sent to a 
frequency counter, 
it might induce incorrect readouts in the frequency 
measurement. 
To notice such a frequency miscount, 
the Ti:S frequency comb at the remote site is 
phase-locked to the SHG light, and 
the in-loop beat signal between the transmitted signal 
and the Ti:S frequency comb is 
counted to confirm 
a stable phase lock. 
Note that the Ti:S frequency comb was sometimes 
phase-unlocked due to the 
intensity variation of the SHG light before installation 
of the polarization control system. 
Thanks to our robust system, the stable intensity 
of the final beat signal between the Ti:S frequency 
comb and the clock laser at the remote site helps
the frequency counter to read out 
infallible and reliable values.

\subsubsection{Evaluation of instabilities induced by other components}

To connect an optical transition in the visible light 
without a performance degradation, 
the transfer stabilities of components to bridge the clock
lasers and 1.5 $\mu$m to be transferred should be lower than the 
fractional instability of the optical clock. The most critical 
component is the EDFA. The length variation of the fiber components 
constituting the EDFA or the amplified spontaneous
emission (ASE) noise causes additional phase noise. 
In our system, a booster EDFA and preamplifier EDFA are used 
before and after the optical carrier transfer system 
to compensate for the large optical loss in the urban link. 
Their performances were evaluated by 
independent measurements using a 
fiber interferometer 
that had arms with and without the device under test, 
and the relative phase instability was measured using the 
heterodyne beat. 
Since the preamplifier EDFA was coupled with the
PPLN, the fiber-pigtailed PPLN was also evaluated together with the EDFA. 
The results of the two instability 
measurements with a $\Lambda$-type counter 
are depicted in Fig. \ref{fig:stability} 
as curves (a) and (b). 
The stabilities of the out-of-loop signal 
and the system noise of the 
interferometer are depicted as curves (c) and (e), 
respectively. 
Since the resultant instabilities of (a) and (b) were in similar level, 
we concluded that the SHG process with the PPLN did 
not degrade the frequency stability. 
Another concern was the phase noise originating 
from the polarization
tracker; however, the results showed that 
our active polarization control 
system neither degraded the frequency stability 
nor gave any frequency offset. 
In addition, the Ti:S frequency combs at both sites and 
the ECDL at the local
site are tightly phase-locked to the reference light,
 and their influences are negligibly small. 
Taking into account all relevant fractional instabilities, 
the overall instability of the all-optical link system 
can be expressed as the root sum square of 
curves (a), (b) and (c) (curve (d)).

\section{Demonstration of direct comparison of two optical clocks}

\begin{figure}[hbt]
\begin{flushleft}
\vspace*{-15mm}
\hspace*{-15mm}
\includegraphics[width=16cm]{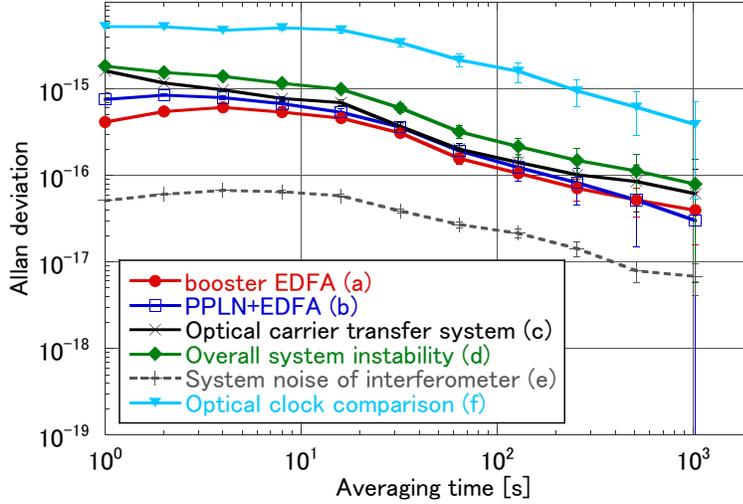}
\vspace*{-38mm}
\caption{Frequency instabilities of direct clock comparison and system
 as measured by a $\Lambda$-type counter. The instability of
 optical clock comparison (f) is not limited by the overall instability
 (d) of the all-optical link.}
\label{fig:stability}
\end{flushleft}
\end{figure}

We tested our novel all-optical link system for making 
a direct comparison of distant optical clocks through a 
60-km urban fiber link in
Tokyo. 
The \(^{87}\)Sr lattice clocks developed at NICT and UT 
were compared. 
The details of the clocks and the Ti:S frequency comb 
can be found in 
\cite{Takamoto2005,Akatsuka2008,Takamoto2011,Yamaguchi2011,Nagano2009}. 
The frequency information was sent from NICT to UT, 
and the fiber noise compensation was done at NICT. 
The beat note representing the differential clock frequency was measured at UT. 
The total stability, including those of the clocks and the overall system, is depicted in curve (f) of 
Fig. \ref{fig:stability}. 
The frequency measurement was done with a $\Lambda$-type 
counter around midnight. 
The short term stability of \(5\times10^{-15}\) at 1 second 
was dominated by the 698-nm clock laser at NICT. 
Fig. \ref{fig:stability} indicates that the 
all-optical link system 
does not put any restrictions on the frequency comparison. 
The system with the 60-km urban fiber link 
enabled us to determine the relative stability of optical
clocks that were 24-km distant from each other. 
Moreover, after correction of the uncommon systematic 
shifts, 
the difference shrank to less than 0.1 Hz with an 
uncertainty of 0.3 Hz attributed to 
the lattice clocks, not the fiber link \cite{Yamaguchi2011}. 
Additionally, a Hz-level frequency difference between 
the clocks is clearly visible 
over the time scale of minutes.

\section{Conclusion}

We established an all-optical link connecting 
two distant optical clocks. 
A 1.5-$\mu$m light was 
stably transferred 
through a 90-km urban fiber link by using 
an optical carrier transfer
system. The transfer stability was 
$2\times10^{-15}$ at 1 second and
$4\times10^{-18}$ at 1000 seconds. 
A polarization control system for the transmitted
light was developed to stabilize the intensity of 
second harmonic light, enabling 
long-term and reliable measurements.
The overall system instability, including the 
instabilities of two EDFAs out of the phase-noise 
compensated path, was 
$2\times10^{-15}$ at 1 second and $7\times10^{-17}$ 
at 1000 seconds. 
The direct comparison of the two $^{87}$Sr lattice 
clocks was realized in the $10^{-16}$ uncertainty 
with the system developed here. 
The instability of the comparison was not limited by the 
overall instability of the all-optical link. 

The short-term stability of optical clocks has recently 
improved to the $10^{-16}$ level \cite{Jiang2011}. 
When such stable optical clocks are to be compared, 
we should use a tracking laser working as low-noise high
gain amplifier instead of the uni-directional EDFA 
employed in the current setup. 
Moreover, a noise-less optical fiber link would facilitate 
more stable transfers.

\section*{Acknowledgments}

The authors would like to thank Tetsushi Takano, Masao Takamoto and
Hidetoshi Katori at the University of Tokyo for 
their operation of the \(^{87}\)Sr clock and fruitful cooperation. 
The authors would like to thank Kazuhiko Nakamura at NICT for his
support on the usage of the JGN2plus optical fiber link as well as 
Ying Li and Clayton Locke at NICT for their cooperation.  
This work was supported by the JSPS through its FIRST program.  

\end{document}